# v-Relax: Virtual Footbath Experiencing by Airflow and Thermal Presentation

Vibol Yem[1,3], Mattia Quartana[2], Zi Xin[3], Kazuhiro Fujitsuka[4], and Tomohiro Amemiya[1]

[1] The University of Tokyo, Hongo, Tokyo, Japan

[2] Tallinn University, Narva, Tallin, Estonia

[3] University of Tsukuba, Tsukuba, Ibaraki, Japan

[4] Mitsui Fudosan Co., Ltd, Chuo, Tokyo, Japan

(Email: yem@iit.tsukuba.ac.jp)

**Abstract ---** Relaxation is a critical counterbalance to the demands of modern business life. Footbaths, a simple yet highly effective therapeutic practice, have been used for centuries across various cultures to promote relaxation and overall well-being. This study presents a novel approach to simulating the experience of a public footbath using tactile and thermal stimulation of airflow to the calf and those on the foot soles. Our system aims to offer a realistic and immersive virtual footbath experience without the need for actual water, by controlling the temperature and airflow to mimic the sensation of soaking feet in water or a water wave. Without using actual water, our system can be more compact, highly responsive, and more reproducible. The layer of airflow is made as thin as possible by adjusting air outlet, and the Coanda effect is also considered to generate a water surface more realistic. The system can provide multi-sensory experience, including visual and audio feedback of water flow, enhancing the relaxation and therapeutic benefits of a footbath.

**Keywords:** Virtual footbath, airflow, temperature, relaxation, virtual reality

## 1 Introduction

Footbaths have long been recognized for their relaxing and therapeutic benefits, aiding in stress relief and improving overall wellbeing. Footbaths with warm water can be used as an easy, simple, and safe nursing intervention to improve sleep quality [1]. However, access to a real footbath may not always be feasible due to time, space, or resource constraints. Moreover, actual water would give a low responsiveness and low reproducibility; thus, it can provide only a limited range of experience.

This study aims to provide public footbath experience using virtual reality (VR) technologies. We explore the potential of using airflow and thermal presentation to recreate the soothing experience. We focus on two main perceptual factors during footbath: the thin layer of water-surface surrounding the foot and water temperature, which are crucial during soaking. Airflow simulation has been focused on presenting tactile sensation feedback [2][3]. However, none of these have been applied to water surface or water wave presentation.

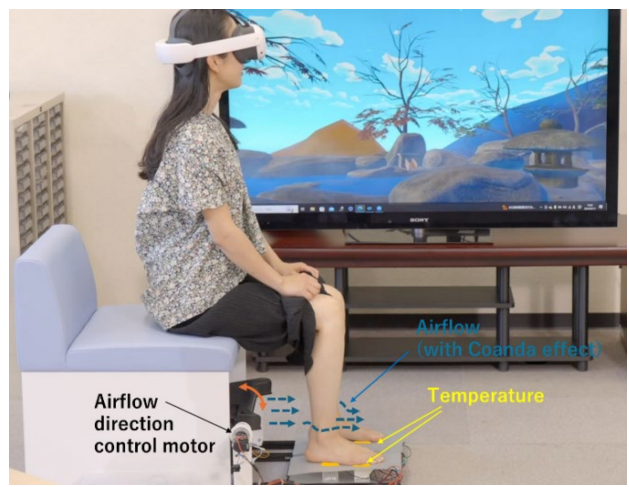

Fig.1  Virtual footbath experiencing

## 2 System

### 2.1 Airflow Simulation and Water Surface Perception

To generate a thin layer of airflow, we designed and developed a narrow air outlet using PLA 3D printing. The air outlet is rectangular in shape, with a width of 1 mm and a total length of 400 mm. To minimize the number of fans, we considered using the Coanda effect to provide a layer of airflow that contacts the skin around the foot (Fig. 1). The Coanda effect is a phenomenon that occurs when a fluid jet, such as air or liquid, adheres to a nearby surface, especially a curved or convex surface. The fluid will follow the shape of the surface, even curving away from its original direction. We considered that the curve of the foot's surface could generate this effect. Therefore, we designed the system to use only one fan (109BG12HC1, Sanyo Denki) to provide the water surface sensation to both feet. Moreover, the air outlet can be moved up or down by a DC motor (203119, Maxon) to generate water wave stimulation while relaxing in the bath.

### 2.2 Heaters and Coolers

A heater film and a temperature sensor (ADT7410, Analog Devices) were placed inside the air box between the fan and the air outlet. The sensor is used for air temperature feedback control. The heater film is turned on for a hotter water sensation, but it is turned off for a cooler sensation. We do not use any cooling device because the user can perceive a cool sensation via the airflow.

Eight Peltier devices (four units for each foot) were used to provide hot/cool sensations to the soles. Temperature sensors were placed next to the Peltier devices for feedback control. For safety, we used two metal plates to cover the Peltier devices and transmit temperature to the sole. Several studies have focused on providing temperature to the sole [4] but providing a water temperature-like sensation is challenging, and our design aims to contribute to this field.

### 2.3 Visual and Audio Sensation Feedback

Visual and audio cues are combined in the virtual environment to create an immersive experience for the users. Users can explore a simulated public footbath environment, visually engaging with their surroundings. Additionally, the system produces sounds such as waterfalls, soaking water, and birds chirping, enhancing the overall sensory experience.

### 3 CONCLUSION AND DEMONSTRATION

The virtual footbath system introduced in this paper showcases an innovative approach to experiencing the soothing effects of a footbath without the use of actual water. By leveraging airflow and precise temperature control, the system provides an immersive and effective relaxation alternative suitable for a variety of environments.

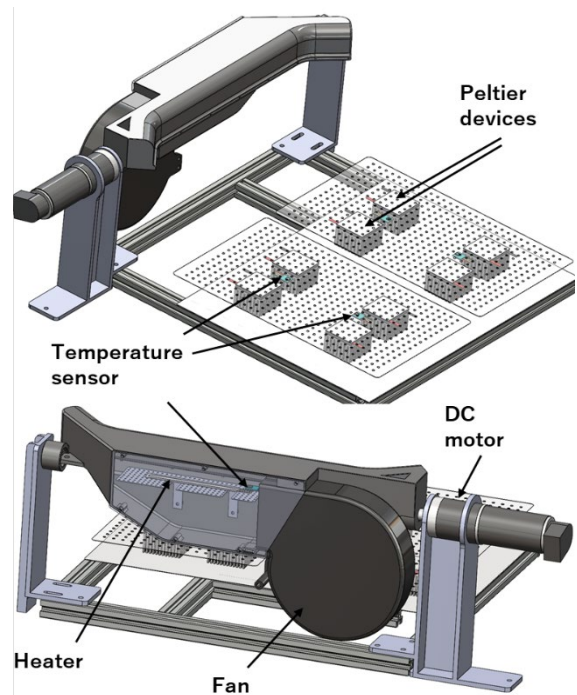

Fig.2 A design of footbath sensation feedback device

During the demonstration, an attendee sits on a chair, places their feet on the device, wears an HMD, and experiences sensory feedback while exploring the virtual environment. For future work, we aim to enhance the system to accommodate multiple users, allowing them to enjoy and interact with one another during experiencing.

ACKNOWLEDGEMENT

This study was financially supported by Mitsui Fudosan Co., Ltd.

REFERENCES

[1] K. Nasiri et al., "The effect of foot bath on sleep quality in the elderly," a systematic review, *National Library of Medicine, MBC Geriatrics*, Vol. 24, No. 1911 (2024)

[2] K. Shimizu et al., "FiveStar VR: shareable travel experience through multisensory stimulation to the whole body," *ACM SIGGRAPH Asia 2018 Virtual & Augmented Reality*, Article No. 2, (2018)

[3] K. Ito, Y. Ban, S. Warisawa: "AlteredWind: Manipulating Perceived Direction of the Wind by Cross-Modal presentation of Visual, Audio and Wind Stimuli." *ACM SIGGRAPH Asia 2029 E-tech* (2019)

[4] A. Mazursky et al.: "ThermalGrasp: Enabling Thermal Feedback even while Grasping and Walking," *IEEE VR 2024*, (2024)